\documentclass{webofc}

\usepackage[varg]{txfonts}   
\usepackage{hyperref}
\usepackage{url}
\hypersetup{colorlinks=true,citecolor=blue,urlcolor=blue,linkcolor=blue}
\graphicspath{ {./plots/} }
\usepackage{xspace}

\newlength{\figwidth}
\newlength{\twofigwidth}
\begin{document}
\title{Longitudinal structure optimization
       for the high density electromagnetic calorimeter}
%
%

\author{\firstname{Oleksandr} \lastname{Borysov}\inst{1} \and
        \firstname{Shan} \lastname{Huang}\inst{2} \and
        \firstname{Kamil} \lastname{Zembaczyński}\inst{3} \and
        \firstname{Aleksander Filip} \lastname{Żarnecki}\inst{3}\fnsep\thanks{\email{filip.zarnecki@fuw.edu.pl}}
}

\institute{Weizmann Institute of Science, Rehovot, 7610001, Israel
     \and
     IFIC, CSIC and Universitat de Val\`encia, C/ Catedr\`atic Jos\'e
     Beltr\'an Mart\'inez 2, 46980 Paterna, Spain
     \and
      Faculty of Physics, University of Warsaw, Pasteura 5, 02-093 Warsaw, Poland}

\abstract{
High density electromagnetic sandwich calorimeters with high readout
granularity are considered for many future collider and fix-target
experiments. Optimization of the calorimeter structure from the point
of view of the electromagnetic shower energy, position and direction
measurement is one of the key aspects of the design. However, mostly
uniform sampling structures were considered so far. We developed a
semi-analytical approach to study calorimeter performance based on the
detailed Geant 4 simulation, which also allows to compare the expected
performance for different non-uniform longitudinal readout
structures. For multi-objective optimization, procedure based on the
genetic algorithm is complemented with non dominated sorting
algorithm. This methodology opens new prospects for calorimeter design
optimization directly addressing specific measurement scenarios or optimization goals. 
}

\maketitle

\section{Introduction}

\begin{figure}[tb]
\centering
  \includegraphics[width=1.0\twofigwidth]{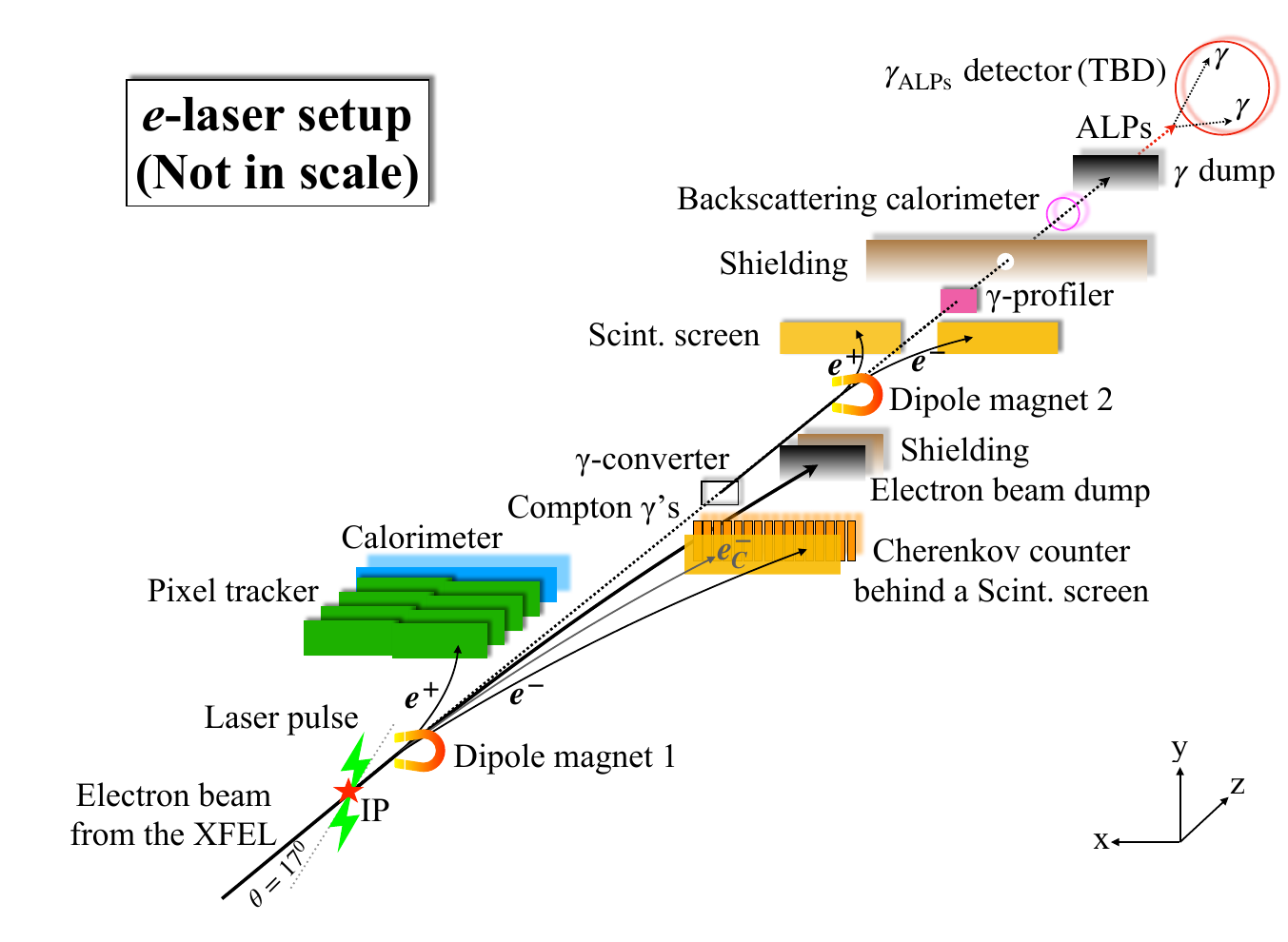}
  \includegraphics[width=1.0\twofigwidth]{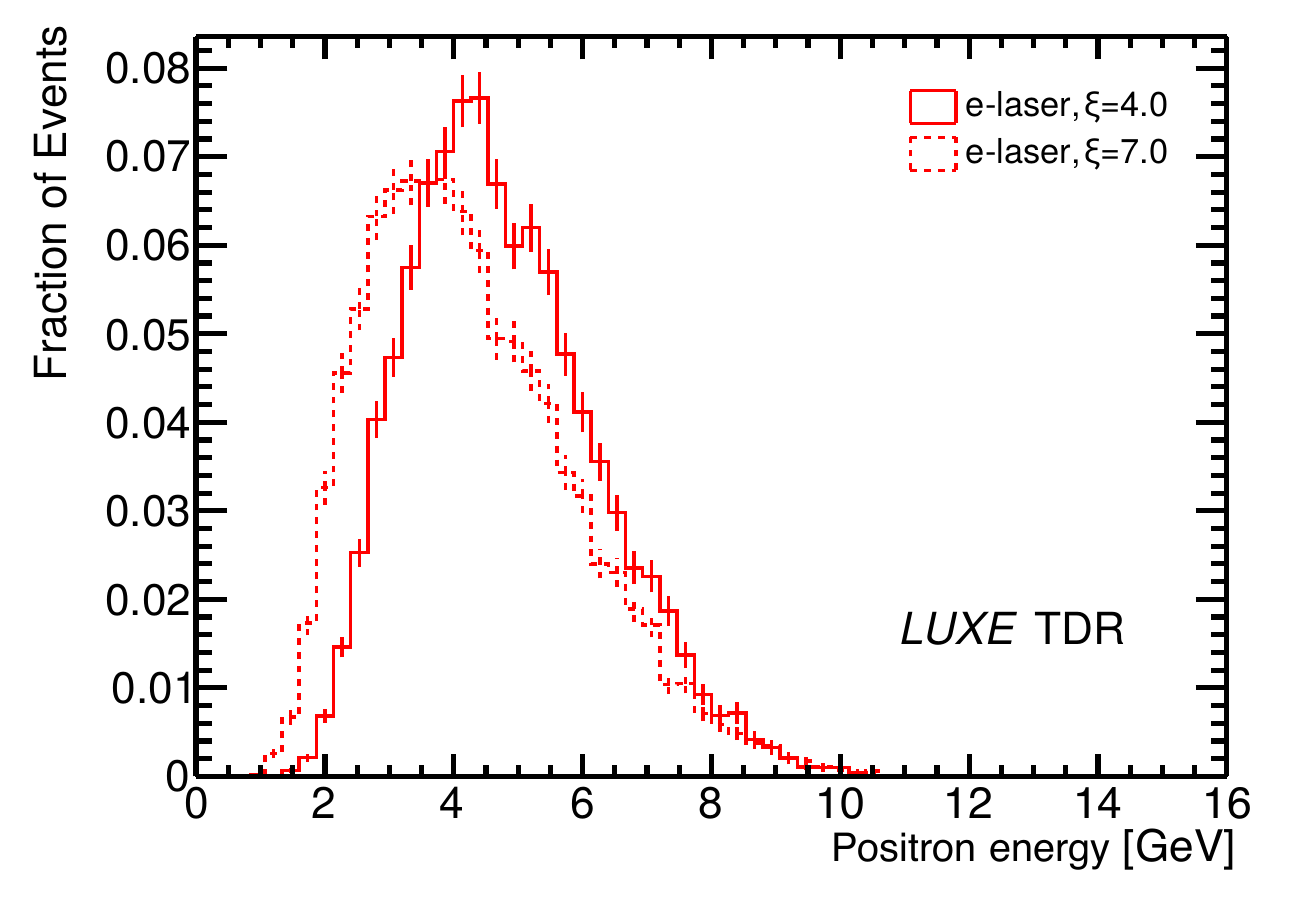}
\caption{Left: schematic layouts for the $e$-laser LUXE setup;
           shown are the magnets, detectors,
           main shielding and beam dump absorbers \cite{LUXE:2023crk}.
           Right: positron energy spectrum for the $e$-laser
            configuration for the initial LUXE phase, for two different values
            of the dimensionless laser intensity parameter $\xi$. }
\label{fig:luxe}
\end{figure}

The LUXE experiment \cite{Abramowicz:2019gvx,Abramowicz:2021zja,LUXE:2023crk}
proposed at DESY will make use of the $16.5$\,GeV
electron beam of the European Free Electron Laser (Eu.XFEL)
colliding with an intense optical laser beam
for research in the field of  the Strong Field Quantum Electrodynamics (SFQED).
The scientific goal of the LUXE experiment is to measure the rate
and spectra of $e^+e^-$ pairs produced in $e$-laser and $\gamma$-laser 
modes,
as well as spectra of Compton electrons and photons in the $e$-laser
configuration.
LUXE will operate in a two-arms planar spectrometer configuration equipped with
calorimeters and trackers to measure the energy and position of electrons
and positrons, see figure~\ref{fig:luxe} (left).
In addition,  downstream from the interaction point
dedicated detectors will be installed to measure the energy
and flux of Compton photons.
Additional components allowing for exotic particle searches are also considered.

Precise measurements of the rates and energy spectra of positrons
produced in the non-linear processes, for different laser intensities
and interaction modes, is a key to exploring SFQED domain.
Example positron energy specta for the $e$-laser
configuration in the initial LUXE phase are shown in
figure~\ref{fig:luxe} (right).
This is however a very challenging measurement for the detector design, as
the number of positrons produced per beam crossing will vary by many orders
of magnitude, depending on the interaction mode in the experiment, and on the laser power.
For low and moderate multiplicities one can resolve
individual showers and measure their energy and positions/angles.
However, for high multiplicities with overlapping showers, only the
total number of positrons and the energy spectrum can be reconstructed.
To meet these requirements, a high density sampling calorimeter made of
tungsten plates and thin silicon sensors was selected for the LUXE
positron calorimeter (ECAL-P).

\section{LUXE ECAL-P design}
\label{sec:ecalp}

\begin{figure}[tb]
\centering
  \includegraphics[width=\figwidth]{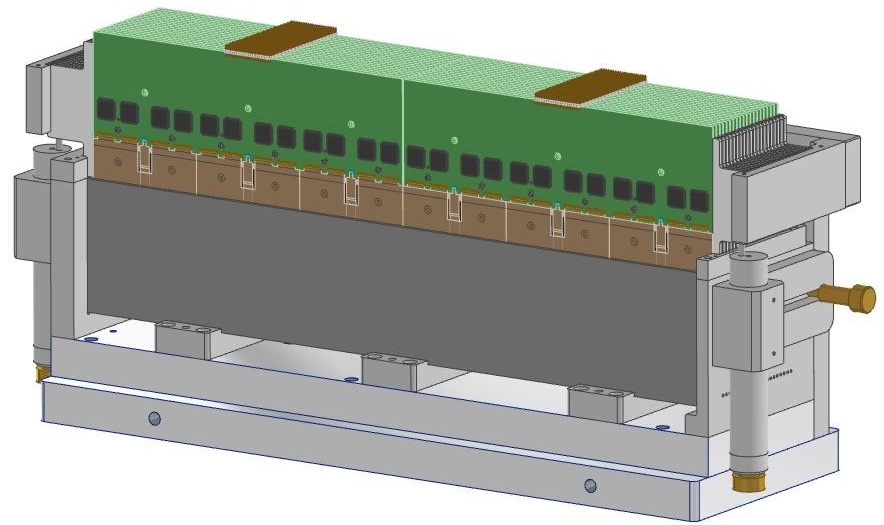}
    \caption{A sketch of the LUXE ECAL-P mechanical structure. The
      frame holds the tungsten absorber plates, interspersed with
      sensor planes. The front-end electronics mounted on the PCBs
      (shown in green) is positioned in the aluminium frames on top of
      ECAL-P and connected with sensor pads via kapton PCB foils~\cite{LUXE:2023crk}.     
    } 
    \label{ECAL}
\end{figure}

The LUXE ECAL-P will consist of $21$ tungsten plates of $1 X_0$
thickness (3.5\,mm) interspersed with $20$ instrumented silicon sensor
planes, placed in a 1\,mm gap between absorber plates~\cite{LUXE:2023crk}.  
The whole structure will be held by an aluminium frame, with slots on
top to position the front-end boards (FEB). A sketch is shown in
figure~\ref{ECAL}.
In the current design, ECAL-P sensors are made of silicon wafers of
$320\,\mu\textrm{m}$ thickness.
Each sensor has a surface of $9 \times 9\, \textrm{cm}^2$ and
consists of 256 pads of $5.5 \times 5.5\,\textrm{mm}^2$ size ($p$ on
$n$-bulk type). Each complete detector plane will consist of six
adjacent sensors. The fiducial volume of the calorimeter will then be
$54 \times 9 \times 9 \, \textrm{cm}^3$.

The ECAL-P will be installed 4.3\,m from the interaction point (IP) on
a special optical table, about 10\,cm behind the silicon tracker. From
simulation and tests of the LumiCal
prototype~\cite{Abramowicz:2018vwb} on which the ECAL-P design is
based, the expected energy resolution is
$\sigma/E=20\%/\sqrt{E/\textrm{GeV}}$ and the position resolution is
about $750\,\mu\textrm{m}$ for electrons of 5\,GeV.
From the deflection in the dipole magnet field, the position
resolution translates into energy resolution of $\sigma/E=0.5\%$ to be compared to 
$\sigma/E=9\%$ from calorimetry measurement alone.
However, for the positron flux and energy spectra reconstruction in the high
multiplicity environment, position and energy measurements from the
ECAL-P need to be combined.
That is why the ECAL-P design has to be optimized for both position
and energy reconstruction precision.

\section{ECAL-P configuration scan}
\label{sec:scan}

The idea of design optimization emerged when studying the expected ECAL-P performance in simulations. 
The question was raised on the possible impact of the reduced number of active sensor planes. 
The number of readout planes could be reduced because of the sensor
damage or readout problems (after sensors and readout electronics are
assembled and installed), or just because of budget constraints.  
This problem could be easily addressed based on the existing Monte
Carlo event samples from Geant4 simulation of the LUXE experiment. 
We focused on the intrinsic energy and position resolution resulting from the
fluctuations in the electromagnetic cascade, while modeling of the silicon
sensor structure and effects of the readout electronics were not taken into account.
Sensor segmentation, signal sampling and amplitude
deconvolution in the readout electronics, threshold and smearing
effects should clearly be taken into account for the final
optimization and performance studies but are not expected to change
the general conclusions.   

\begin{figure}[htb]
\begin{center}
\includegraphics[width=1.0\twofigwidth]{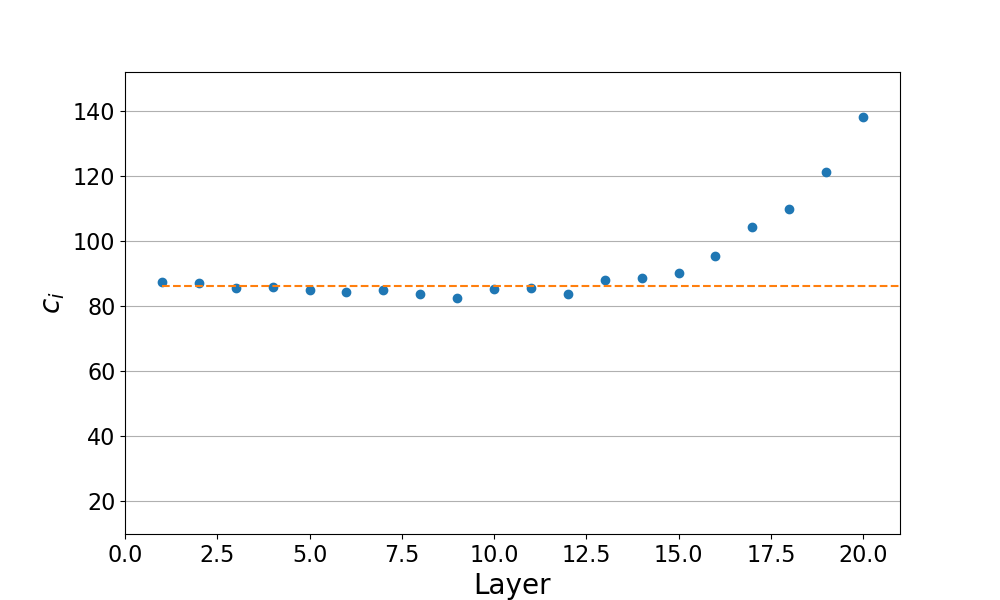}
\includegraphics[width=1.0\twofigwidth]{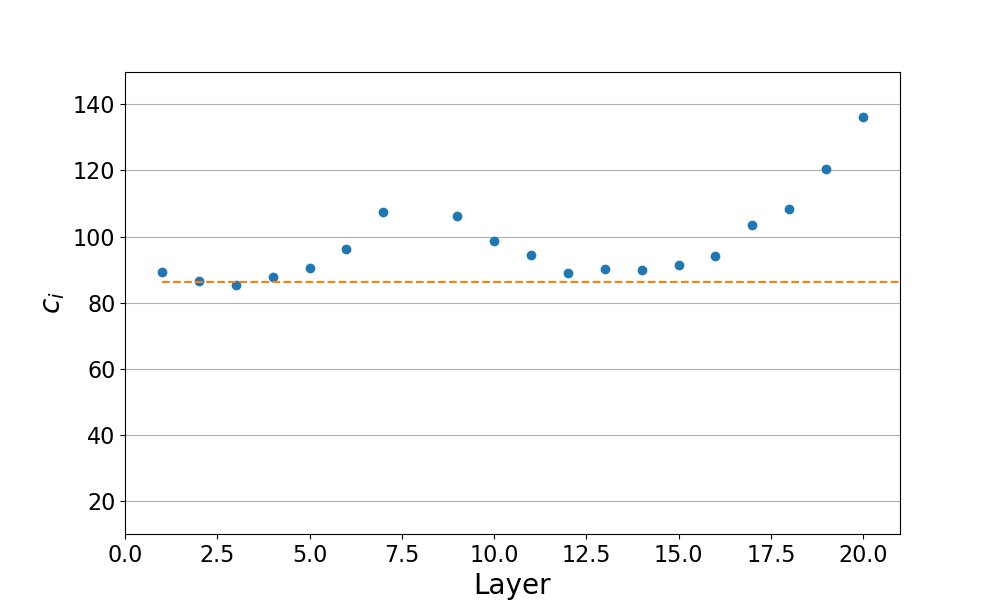}
\end{center}
\caption{Calibration results for the LUXE positron calorimeter, as obtained with Geant4 simulation, for energy resolution optimization in the 2.5 -- 15\,GeV range. Calibration factors of subsequent calorimeter layers are shown for a calorimeter with all sensor layers (left) and for the test scenario with the eighth sensor layer missing (right). The horizontal dashed line indicates the calibration factor value for the uniform calibration of the fully instrumented calorimeter.}
\label{fig:opt_par}
\end{figure}

It was almost immediately realized that for the non-uniform
calorimeter structure (which is the case when some of the sensor
layers are not instrumented or the sensors are broken) recalibration
of the remaining layers is a must. 
When the longitudinal shower leakages can be neglected, distribution
of the calorimeter response for a given positron energy can be very
well described by the Gamma distribution, 
which has two free parameters. 
A convenient choice is to use average calorimeter response and the RMS
of the distribution.  
Calibration factors for the subsequent calorimeter sensor layers can
be then found by solving the minimization problem: looking for the
minimum of RMS with a fixed value of the average calorimeter response
(equal to the initial positron energy).  
An analytical solution to this problem was found, resulting in a very
fast and efficient calibration procedure, which can be applied to many
calorimeter instrumentation scenarios.  
Example results of the proposed calibration procedure are shown in Fig.~\ref{fig:opt_par}.
Calibration factors obtained from the calibration procedure performed
in the 2.5 -- 15\,GeV positron energy range are shown for two
scenarios.  
When all calorimeter layers are instrumented (left plot), a uniform
calibration is obtained for most of the calorimeter layers (up to
layer 15, corresponding to the calorimeter depth of 15 X$_0$) and the
increase of the calibration factors for the last layers can be
understood as an effective way to correct for the longitudinal
leakages for late showers.  
If one of the calorimeter layers is removed from the readout (right
plot), the optimisation procedure results in significant increase of
the calibration factors for the neighbouring layers, in an attempt to
correct (on average) for the signal losses in the nonactive layer.

\begin{figure}[tb]
\centering
\includegraphics[width=1.0\twofigwidth]{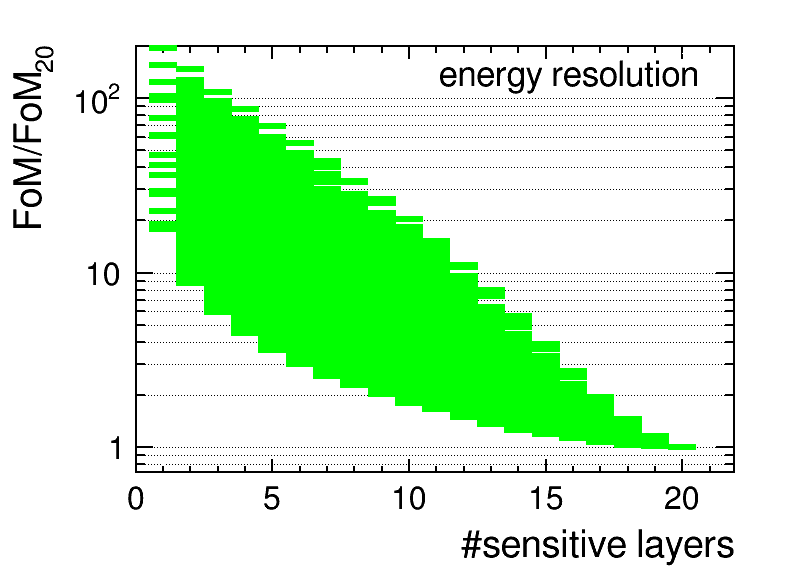}
\includegraphics[width=1.0\twofigwidth]{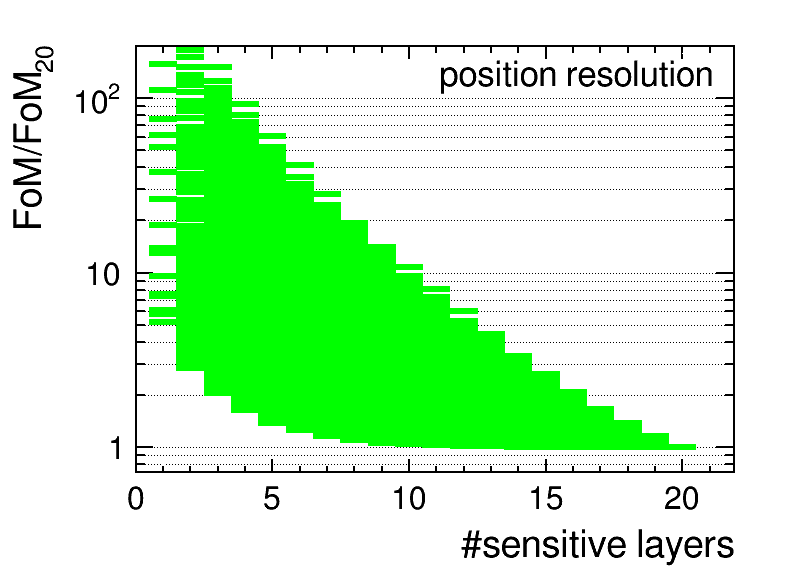}
\caption{Results of the longitudinal readout structure optimization of
  the LUXE positron calorimeter. Figure-of-merit (FoM) for the energy
  resolution (left) and  FoM for the position
  reconstruction (right) are shown as a function of the number of
  instrumented sensor layers, relative to the optimal 20 layer
  configuration result. }
\label{fig:opt_1d}
\end{figure}

With the very fast calibration procedure described above,
comparison of the expected performance for many different calorimeter
instrumentation schemes became possible. In fact, for the assumed LUXE
ECAL-P structure with 20 sensor gaps, all possible instrumentation
scenarios could be tested.   
This is illustrated in Fig.~\ref{fig:opt_1d} (left), where the energy
resolution figure-of-merit (FoM) is shown for all scenarios as a function of
the number of instrumented layers. The optimization was carried out
for positrons perpendicular to the face of the calorimeter.  
It is clear that the reduction in the number of sensor layers results
in worse energy resolution, but the effect of going from 20 to 18
layers is marginal, increasing fast only below 15 instrumented layers. 

Optimal calibration of the calorimeter is the one resulting in the
best energy resolution and best linearity of the response (these are
two different optimization goals, but they are combined in the
developed procedure). 
However, a similar optimization procedure can be applied to the problem
of positron position measurement, when we look for the optimal
weighting factors to calculate the shower position from average
positions in subsequent layers. 
Again, we can then check how removing some of the sensor layers
affects the positron position reconstruction. This is shown in
Fig.~\ref{fig:opt_1d} (right), where the position resolution FoM, defined as the expected resolution of the weighted position
average, is shown as a function of the number of active readout
layers.  
It is clear that the dependence is very different from the one
observed for the energy resolution (left plot).  
As the width of the developing electromagnetic shower increases with
calorimeter depth, the best position reconstruction is obtained from
the first few layers. Position reconstruction in the last layers of
the calorimeter have hardly any effect on the reconstructed initial
positron position.  
Even with only 10 active layers, we can find a configuration resulting
in position reconstruction precision only 5\% worse than with 20
layers. 

\begin{figure}[tb]
\centering
\includegraphics[width=1.0\figwidth]{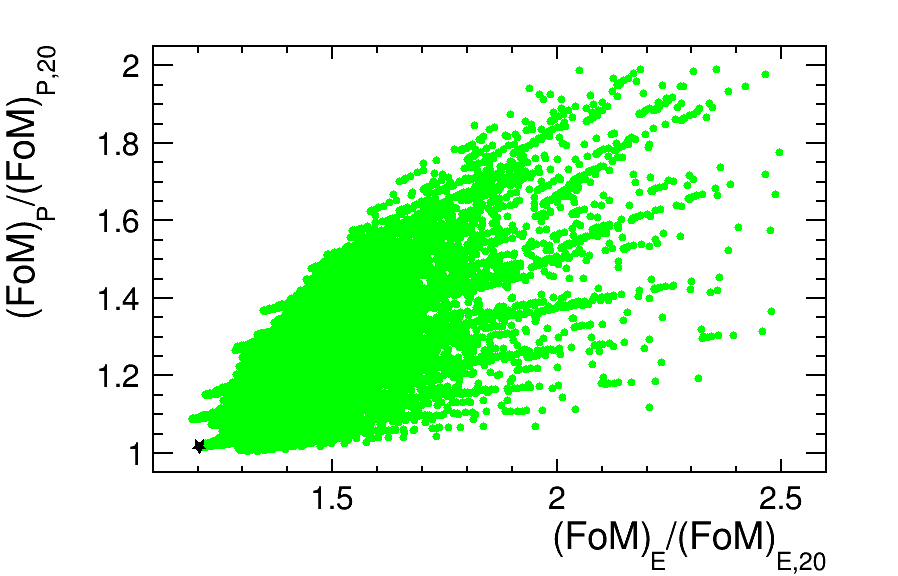}
\caption{Results of the longitudinal readout structure optimization of
  the LUXE positron calorimeter. The FoM for the position
  resolution, (FoM)$_\textrm{P}$,  as a function of the FoM for the energy resolution, (FoM)$_\textrm{E}$, for different configurations of 15
  instrumented layers, relative to the 20 layer configuration result.} 
\label{fig:opt_2d}
\end{figure}

When looking for the optimal detector design, all relevant aspects of
its performance (like response linearity, energy resolution, shower
position and angle measurement precisions) should be taken into
account.  
In most cases, different optimization goals will result in different
design options being selected. 
This is illustrated in Fig.~\ref{fig:opt_2d}, showing the distribution
of the position resolution and energy resolution FoM for
all calorimeter configurations involving 15 active layers. 
Although there is some correlation between the two parameters, the
best energy resolution and the best position resolution cannot be
achieved at the same time; they correspond to different
instrumentation schemes. 
However, the scenario resulting in almost optimal energy resolution
and almost optimal position resolution at the same time, can be easily
selected in the presented case.
This scenario, with calorimeter gaps 2 to 14, 16 and 17 instrumented
with silicon sensors, is indicated with a black star in
Fig.~\ref{fig:opt_2d}. 

\section{Monte Carlo optimization}
\label{sec:mc}

Optimization of the ECAL-P prototype sensitive layer layout described
above was a relatively simple task. With 20 slots between tungsten
layers, the number of geometry configurations to consider is
relatively small ($10^{6}$). With a fast analytical calibration
procedure (few $\mu$s per configuration) the problem can be solved
directly by checking all of these configurations. 
However, this approach is not possible when we want to consider a more
general optimization problem.

For obtaining optimal energy or position resolution one should allow
for non-uniform longitudinal structure of the calorimeter, with
varying distances between subsequent sensor layers.
However, the most general approach, allowing for absorber layers of arbitrary
thickness is difficult to implement as the new GEANT4 simulation (very
time consuming for large event samples) needs to be run for each
considered configuration.
As a compromise, we created a dedicated ECAL-P model with 75 tungsten
plates  of $\frac{1}{3}$\,X$_0$ each and 75 active layers with
320\,$\mu$m silicon sensors in $\frac{1}{3}$\,mm gaps between them.
With this structure we obtain almost the same average density of the
calorimeter as for the baseline LUXE ECAL-P design, but the
calorimeter is extended to 25\,X$_0$ and the position of sensor layers
can be selected with the  $\frac{1}{3}$X$_0$ precision.

\begin{figure}[tb]
\centering
\includegraphics[width=1.0\twofigwidth]{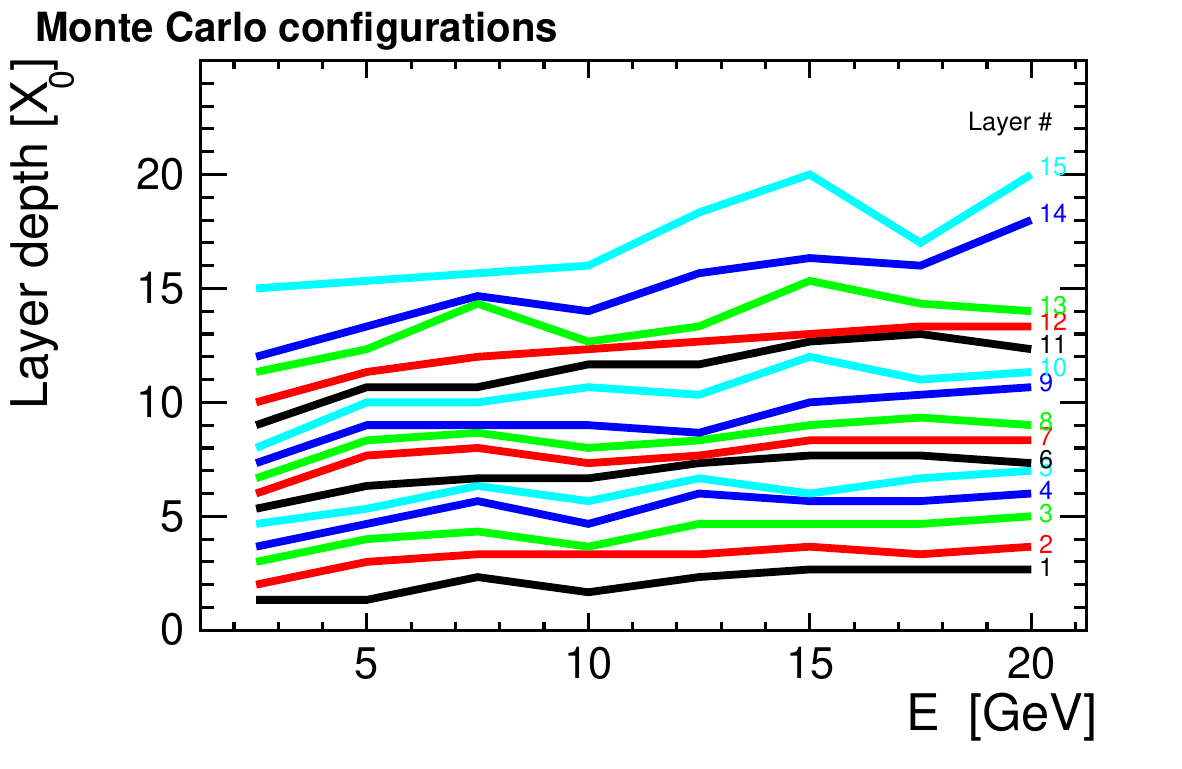}
\includegraphics[width=1.0\twofigwidth]{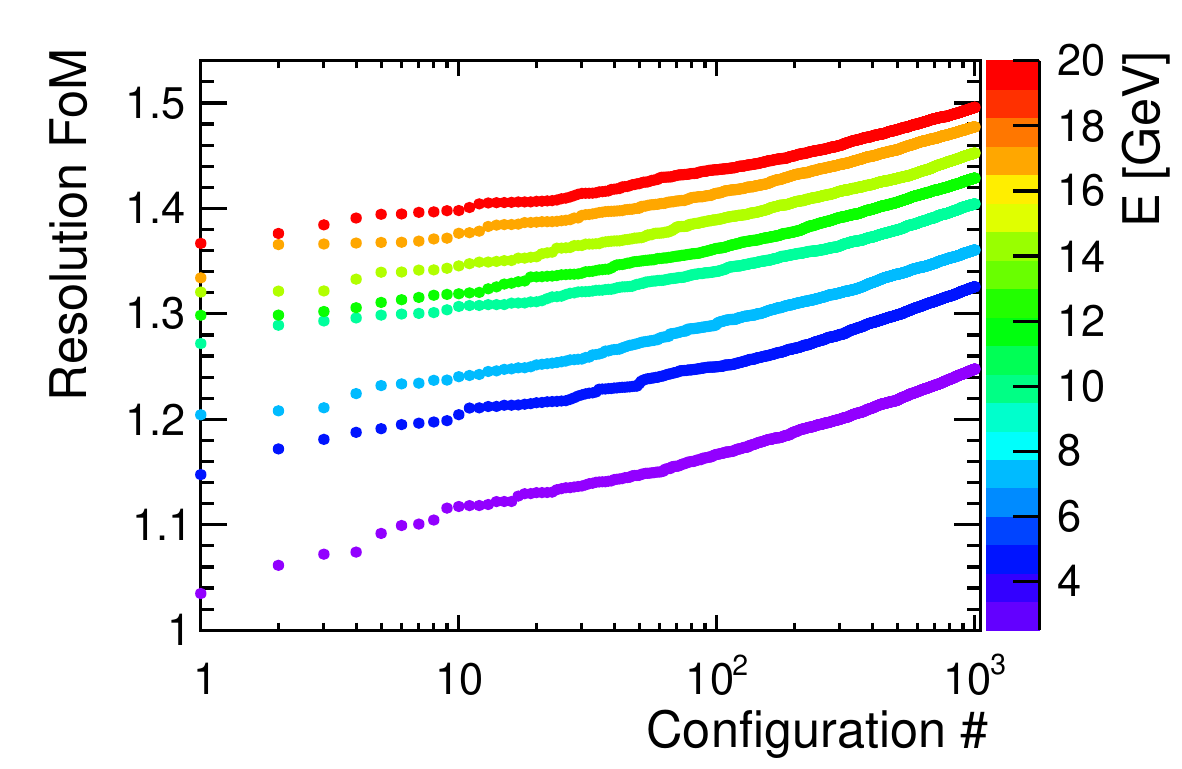}
\caption{Results of the Monte Carlo optimization of the calorimeter
  configuration, for 15 active layers, based on one million random
  layer configurations generated for each energy. Left: best layer
  configuration as a function of energy. Right: resolution FoM for the best 1000 layer configurations at different energies. }
\label{fig:mc}
\end{figure}

With the assumed detector model there are about $2\cdot 10^{15}$
configurations possible for a calorimeter with 15 active sensor layers.
While scanning over all configurations is clearly not possible, we
tried to use Monte Carlo method, selecting configurations at random,
to look for optimal layer configuration.
Example results of such a procedure are shown in figure\,\ref{fig:mc} (left),
where positions of the sensor layers, optimal from the point of view of
the energy resolution, are shown as a function of the positron
energy for a calorimeter with 15 sensor layers
Positions shown are calculated as average layer positions from the
100 configurations with the best energy resolution out of the million
of random configurations generated for a given energy. 
Two effects can be observed. Optimal calorimeter structure depends
on the initial energy: layer depths increase with energy as expected,
reflecting lengthening of the longitudinal profile of the
electromagnetic cascade. But also, in the general approach used,
optimal energy resolution is obtained for a non-uniform longitudinal
structure, with different thicknesses of the subsequent absorber
layers.  
On the other hand, although $10^6$ random configurations were
generated per energy point, significant fluctuations are still visible
in the results, for the last sensor layers in particular.
This is also illustrated in figure\,\ref{fig:mc} (right) where the
energy resolution FoM, defined as the energy averaged
ratio of the expected energy resolution to the reference resolution of
$20\%/\sqrt{E}$, is shown for the best 1000 Monte Carlo configurations
(ordered in resolution) at each energy.
There are significant differences between the subsequent
configurations selected, showing that the Monte Carlo procedure is 
far from converging to the optimal choice.
This shows that the Monte Carlo approach itself is not efficient
enough to find optimal solutions and it has to be complemented by a
second stage, where modifications of the initially selected
configurations should be considered.

\section{Genetic algorithm}
\label{sec:gen}

\begin{figure}[tb]
\centering
\includegraphics[width=0.95\twofigwidth]{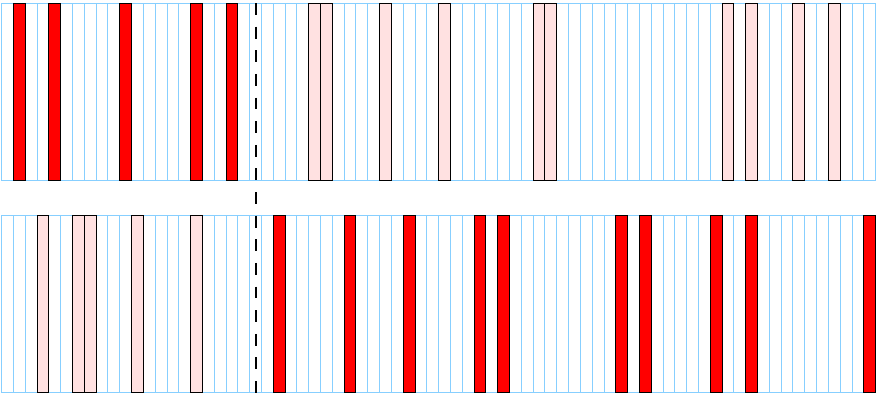}
\hspace*{0.1\twofigwidth}
\includegraphics[width=0.95\twofigwidth]{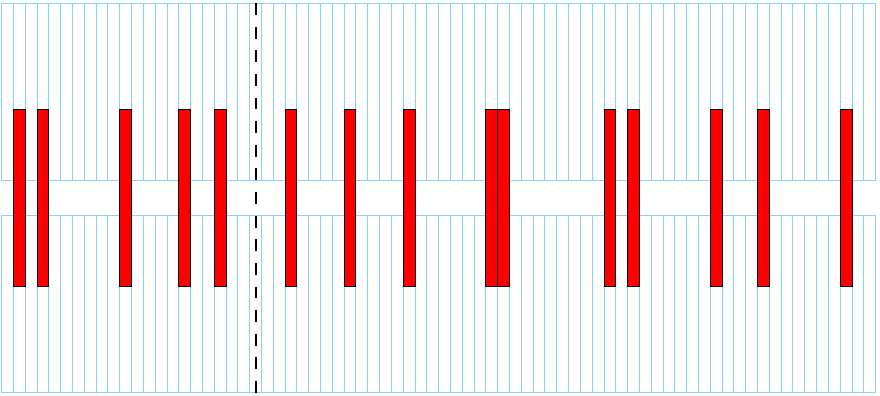}
\caption{Schematic presentation of the genetic algorithm concept.
Left: two parents are randomly selected from the collection of best
configurations and a random cut in the layer sequence is used to
combine the first part of the first parent with the second part of the
second one. Right: random mutations are added to the combined layer
sequence.}
\label{fig:parents}
\end{figure}

The Monte Carlo approach is a convenient way to select the initial sample
of configurations with a reasonable calorimeter performance.
A Genetic algorithm \cite{Mitchell} can be then explored to evolve this
sample towards optimal configurations.
The Genetic algorithms are inspired by biological evolution processes,
such as reproduction, mutation, recombination, and selection.
Implementation of this algorithm to the calorimeter configuration
evolution is illustrated in figure~\ref{fig:parents}.
At each step of the algorithm, we select the 1000 best candidates from the
currently considered configuration-set as potential ``parents'' for the
next generation.
We then generate 10\,000 ``children'' by randomly selecting
two parents and merge their layer positions based on a random cut in
the layer sequence (see left plot in figure~\ref{fig:parents}).
At the last step, random mutations are added to the combined layer
sequence.
Mutation size (change of layer gap index) are generated from
a Poisson distribution (and including random sign) with initial average
mutation size of 1, decreasing with each step of the algorithm.
If the sequence of layer indexes is ``broken'' at any step of the
algorithm (not in strictly increasing order), generation of a given
child is repeated starting from the very beginning (parent selection).

\begin{figure}[tb]
\centering
\includegraphics[width=1.00\twofigwidth]{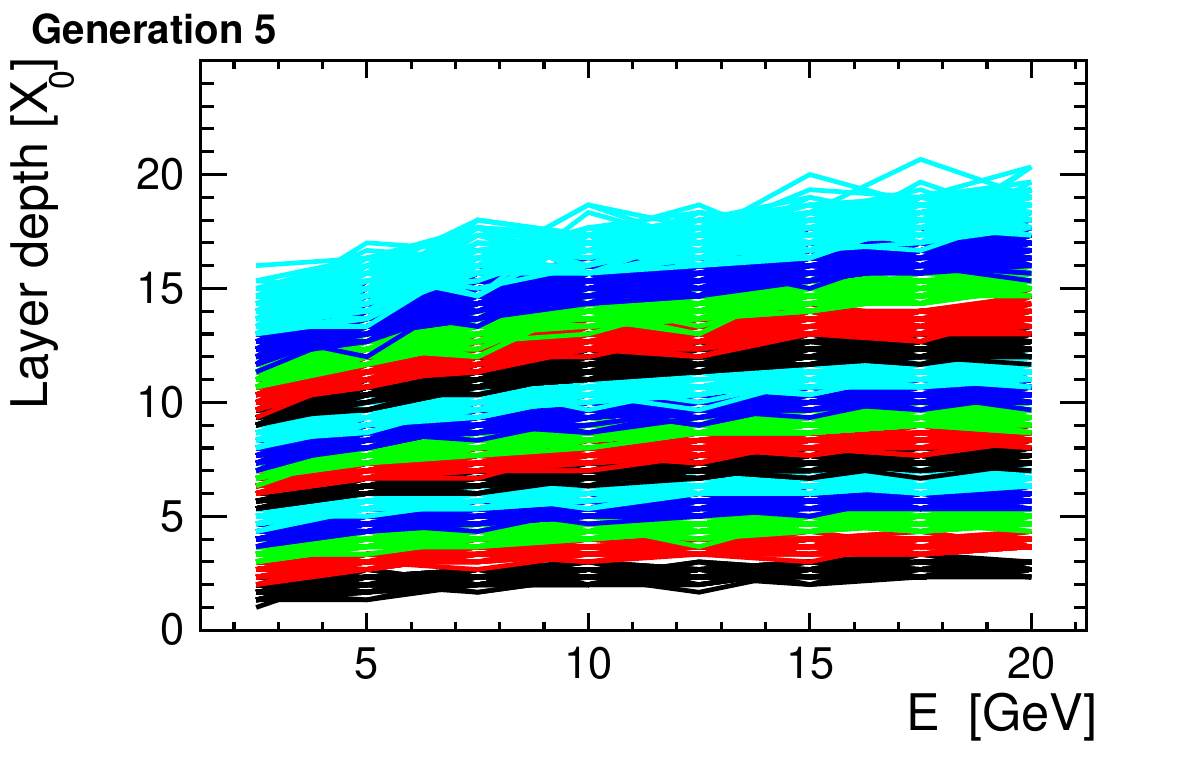}
\includegraphics[width=1.00\twofigwidth]{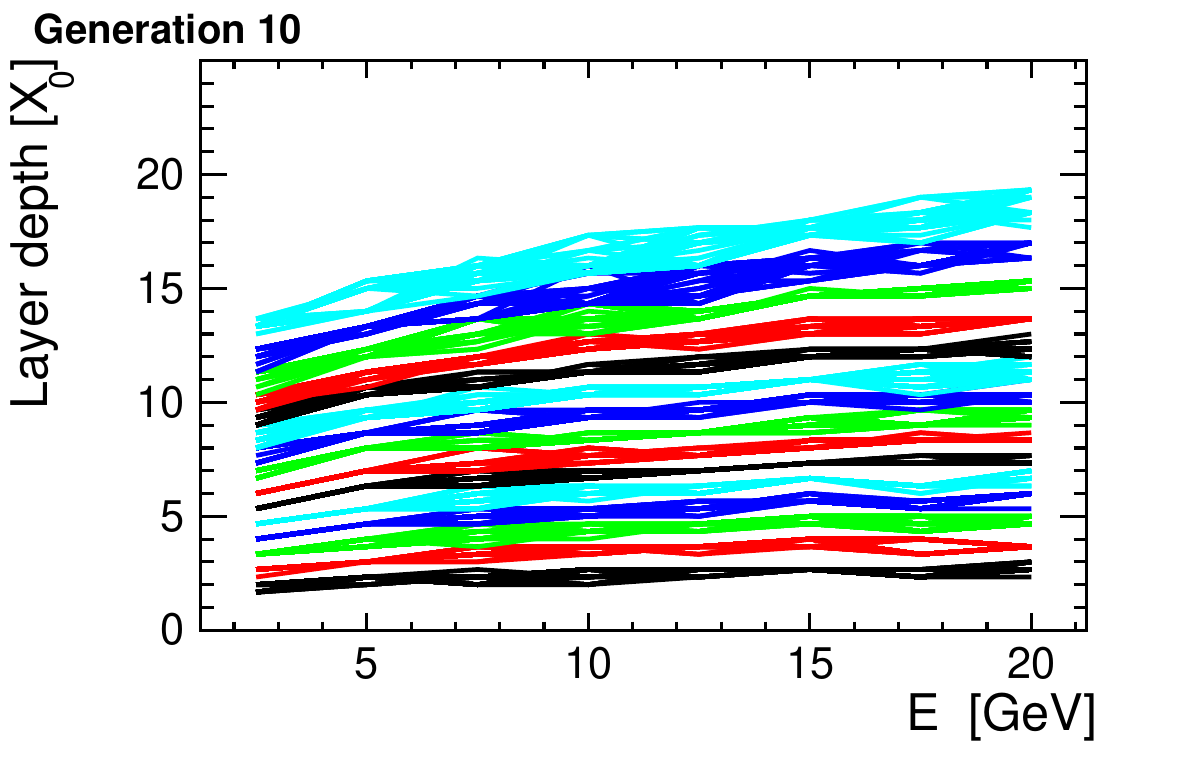}
\includegraphics[width=1.00\twofigwidth]{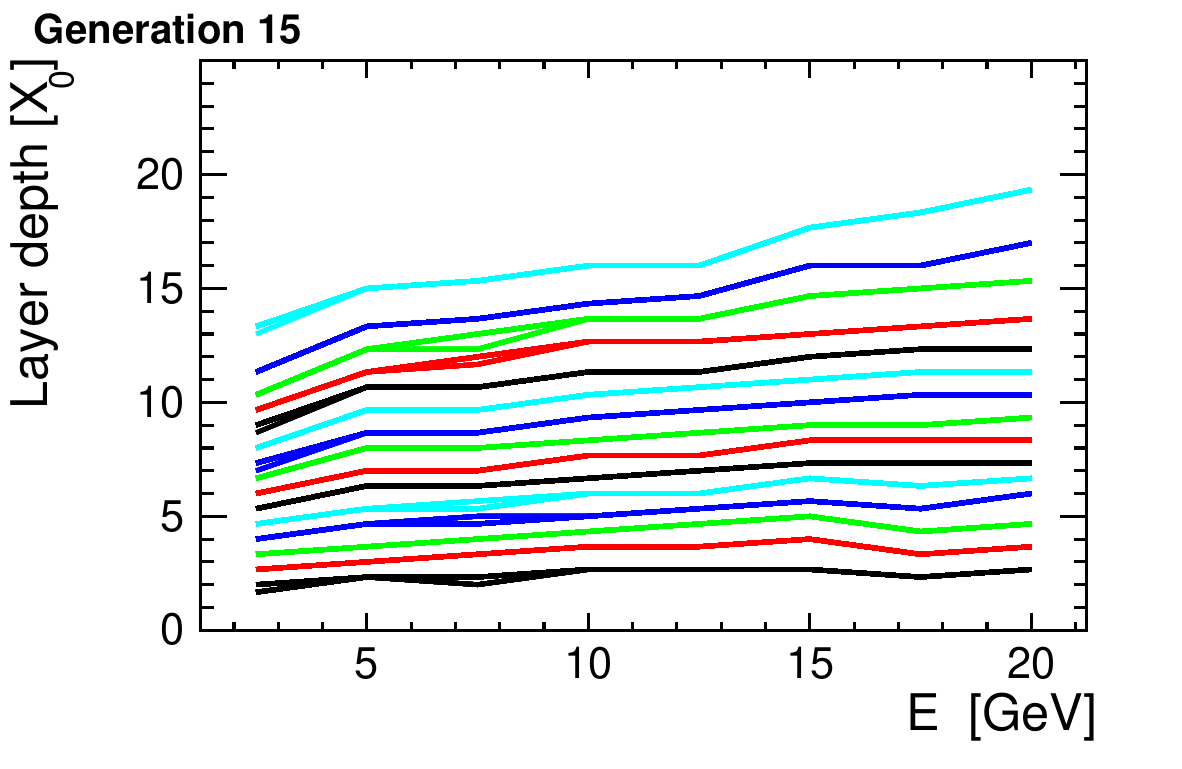}
\includegraphics[width=1.00\twofigwidth]{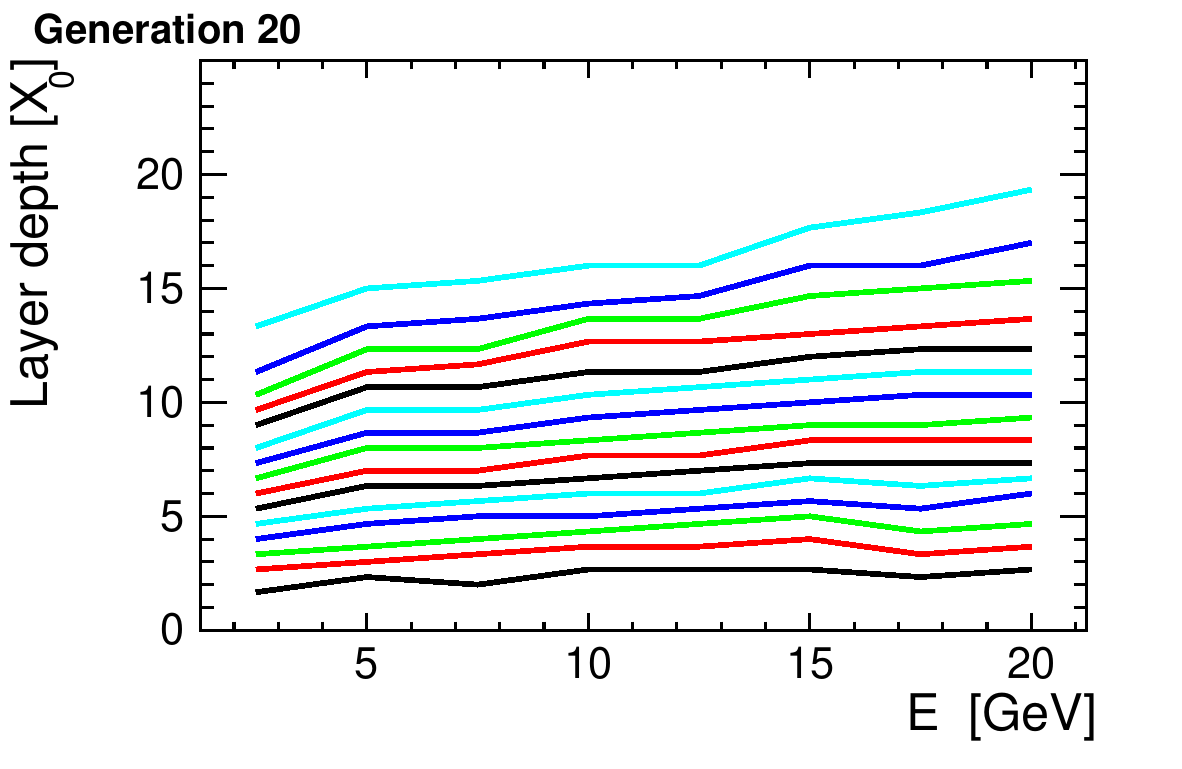}
\caption{Positions of the calorimeter layers as a function of
  energy for 100 configurations with the best energy resolution,
  resulting from the genetic algorithm application with 5,
  10, 15 and 20 generations.} 
\label{fig:genres}
\end{figure}

\begin{figure}[tb]
\centering
\includegraphics[width=\figwidth]{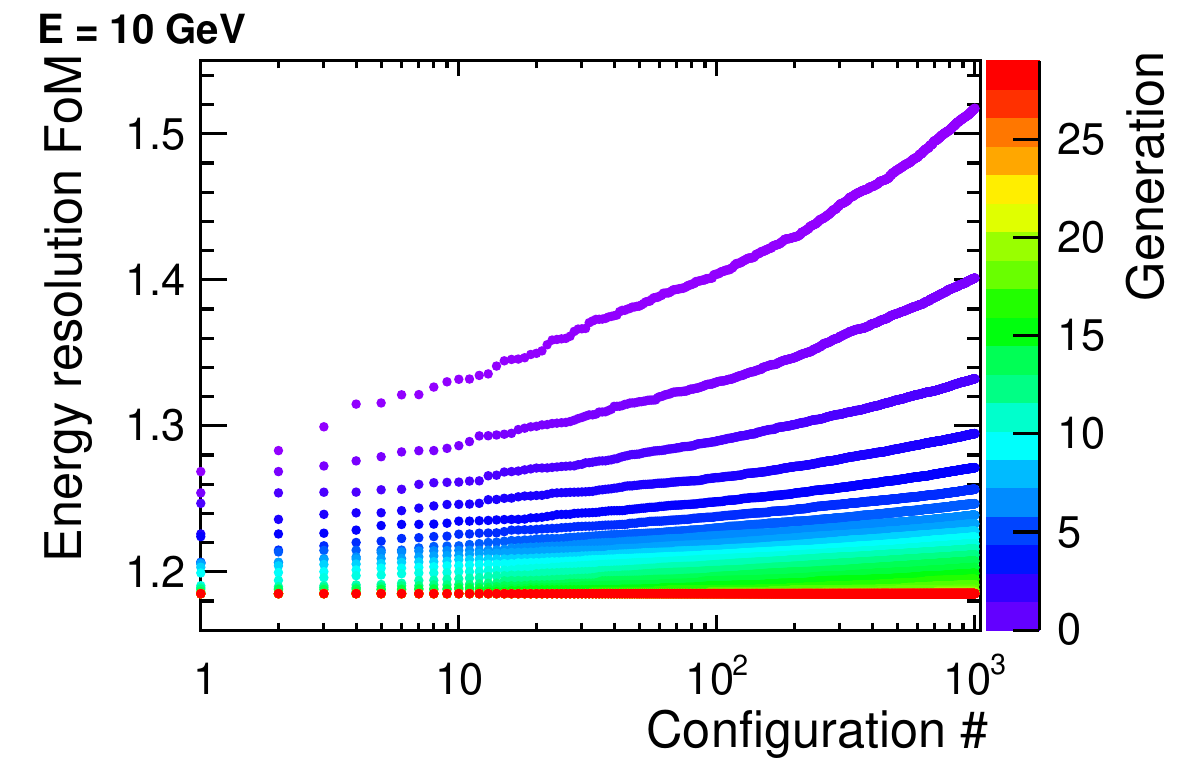}
\caption{Energy resolution FoM for the best 1000 layer
  configurations at 10\,GeV energy and different generations of the
  calorimeter configurations. The first generation of parents is selected
  from a random set of layer configurations. 
} 
\label{fig:genrank}
\end{figure}

Performance of the genetic algorithm for the energy resolution
optimisation problem is presented in figure~\ref{fig:genres}.
Presented are layer positions for 100 configurations with best energy
resolution in each shown generation.
Starting from the random selection of configurations (Monte Carlo
optimization), large differences are observed between the best
configurations generated with genetic algorithm even after 10 steps.
This is also due to sizable mutations introduced in the algorithm
which smear the initial configurations.
Still, after 20 steps of the algorithm, when the mutation level is already
negligible, we end up with a uniform set of configurations (all best
configurations are the same).
This is also shown in figure~\ref{fig:genrank}, where the energy
resolution FoM for 10\,GeV incident positron energy is
shown for the best 1000 configurations (ordered in resolution) from
each generation.
While there are significant differences between best solutions in the
first few generations, the generation becomes almost uniform after 20
generations and no further improvement is possible (the algorithm was
stopped after 30 generations). 

Two effects which were mentioned in section~\ref{sec:mc} when
discussing results shown in figure~\ref{fig:mc}\,(left) are much
better visible in the last plot in figure~\ref{fig:genres}
(configurations after 20 genetic algorithm steps) as the fluctuations
are largely suppressed. Optimal calorimeter layer depths increase with
energy, reflecting lengthening of the longitudinal profile of the
electromagnetic cascade and the optimal energy resolution is obtained
for a non-uniform longitudinal structure, with the largest density of
readout layers in the region of the cascade maximum.

\section{Multi-objective optimization}

\begin{figure}[tb]
\centering
\includegraphics[width=\figwidth]{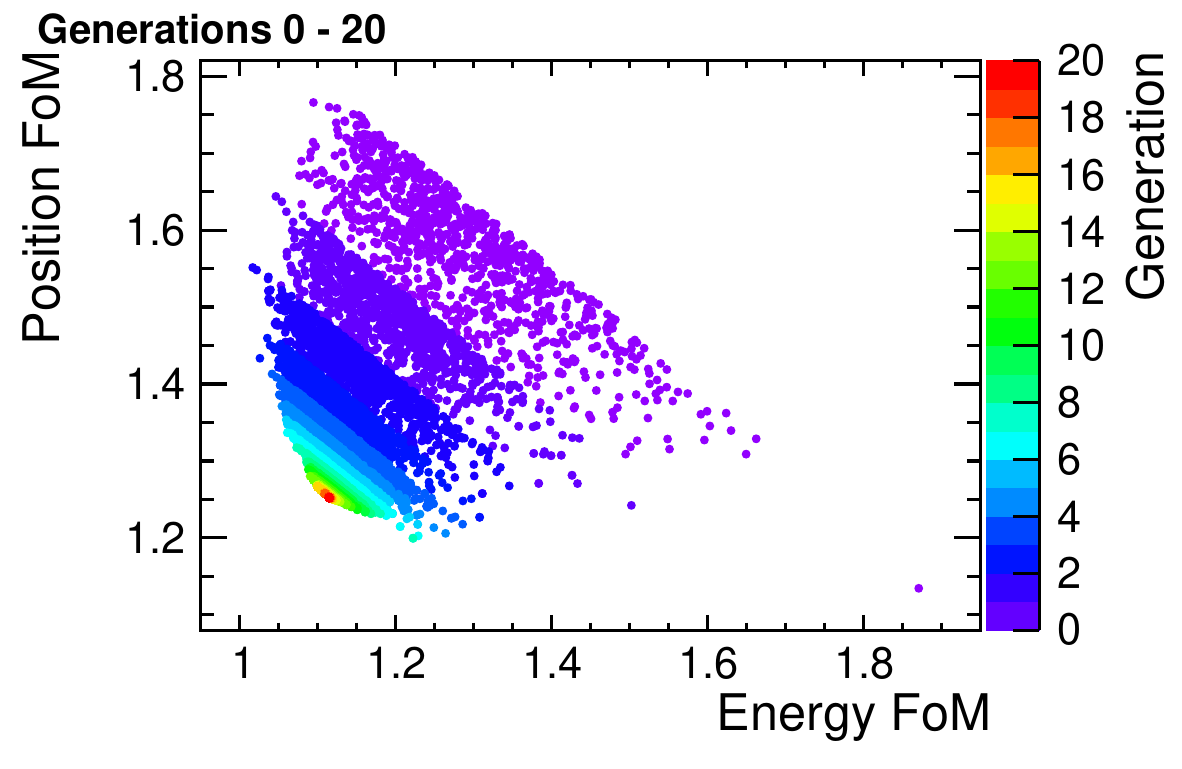}
\caption{Example of the genetic algorithm application to the
  multi-objective problem. The Energy FoM as a function of the position FoM
  for the best 1000 layer configurations from subsequent
  configuration generations. Optimized is the combined FoM, 
  defined as a sum of the energy and position FoMs.
} 
\label{fig:multi}
\end{figure}

Procedures described in sections \ref{sec:mc} and
\ref{sec:gen} focused on energy resolution optimization only. However,
as already discussed in sections \ref{sec:ecalp} and \ref{sec:scan}
the ECAL-P design has to be optimized for both position and energy
reconstruction precision.
If the optimization goal can be uniquely defined in terms of the
measurable performance parameters, the optimization procedure can be 
adjusted accordingly, based on a new figure of merit.
Example of the genetic algorithm application to the multi-objective
problem is presented in figure~\ref{fig:multi}. FoM for
the calorimeter structure optimization is defined as the sum of the
energy resolution and position resolution FoMs. 
As was the case previously, optimization based on the genetic
algorithm converges fast to a single solution.

Unfortunately, a universal FoM can not be defined in most
cases.
For the LUXE ECAL-P, relative importance of position and energy
resolution in the data analysis will depend on the experimental
conditions (type of collisions, electron beam quality, laser beam
intensity etc.) and the measurement goal. 
Also the performance of the silicon tracker, placed in front of the
calorimeter, can have significant impact on the analysis.
In the most general approach, when optimizing the longitudinal structure of the
calorimeter, we should try to look for the procedure allowing for both
energy and position measurement optimization, but without specifying a
particular optimization goal.

We decided to solve this problem with a non dominated sorting procedure.
This idea was previously successfully used in the top-quark
threshold scan optimization at CLIC \cite{Nowak:2021xmp}.
It is based on the observation that, even when we look at two
different objectives, partial sorting of different configurations
should still be possible.
When configuration A gives better energy resolution and better
position resolution than configuration B, we can clearly state that A
is better (more optimal) than B.
However, if only one resolution is better and the other one is worse,
we can not decide which configuration is better (without considering
particular measurement goals). They have to be considered as
equivalent.
By grouping configurations in equivalent sets we form the so-called
Pareto fronts, and we can (at least partially) sort all considered
configurations and select the best performing ones (by selecting best
performing fronts) without any additional assumptions!
This sorting procedure can be used together with a genetic algorithm to
find the best performing calorimeter configurations.

\begin{figure}[tb]
\centering
\includegraphics[width=\twofigwidth]{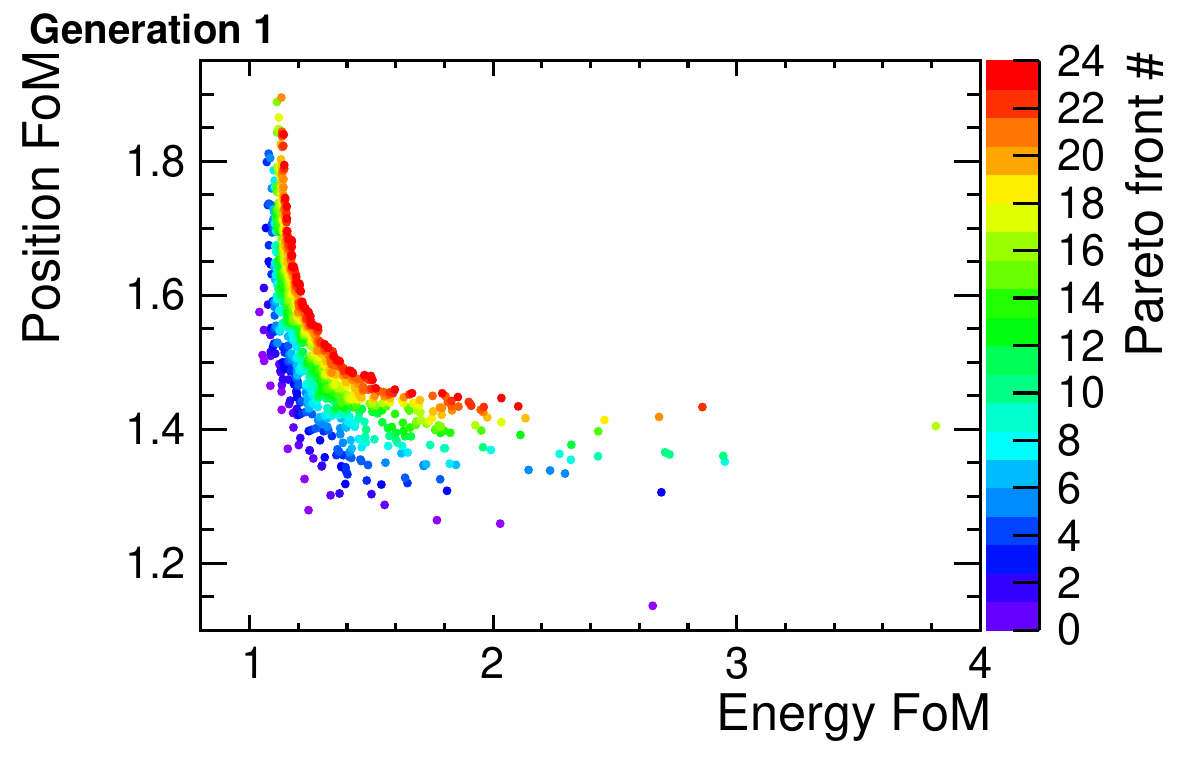}
\includegraphics[width=\twofigwidth]{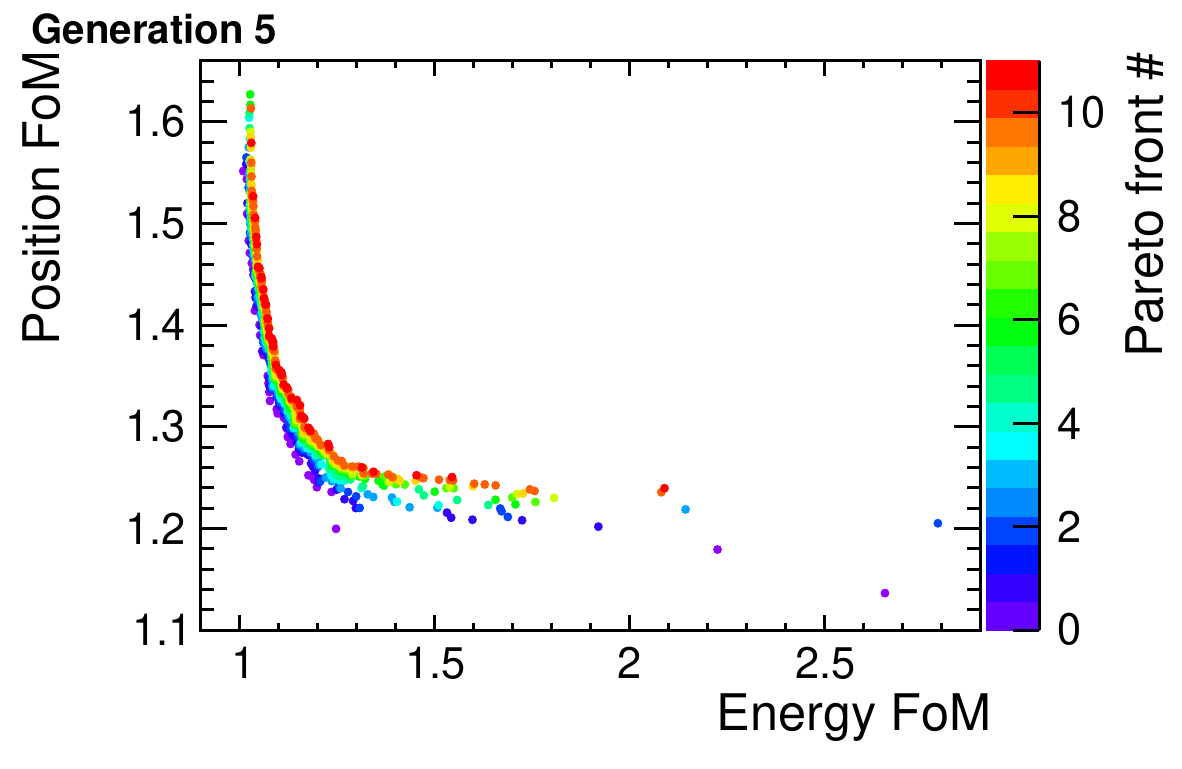}
\includegraphics[width=\twofigwidth]{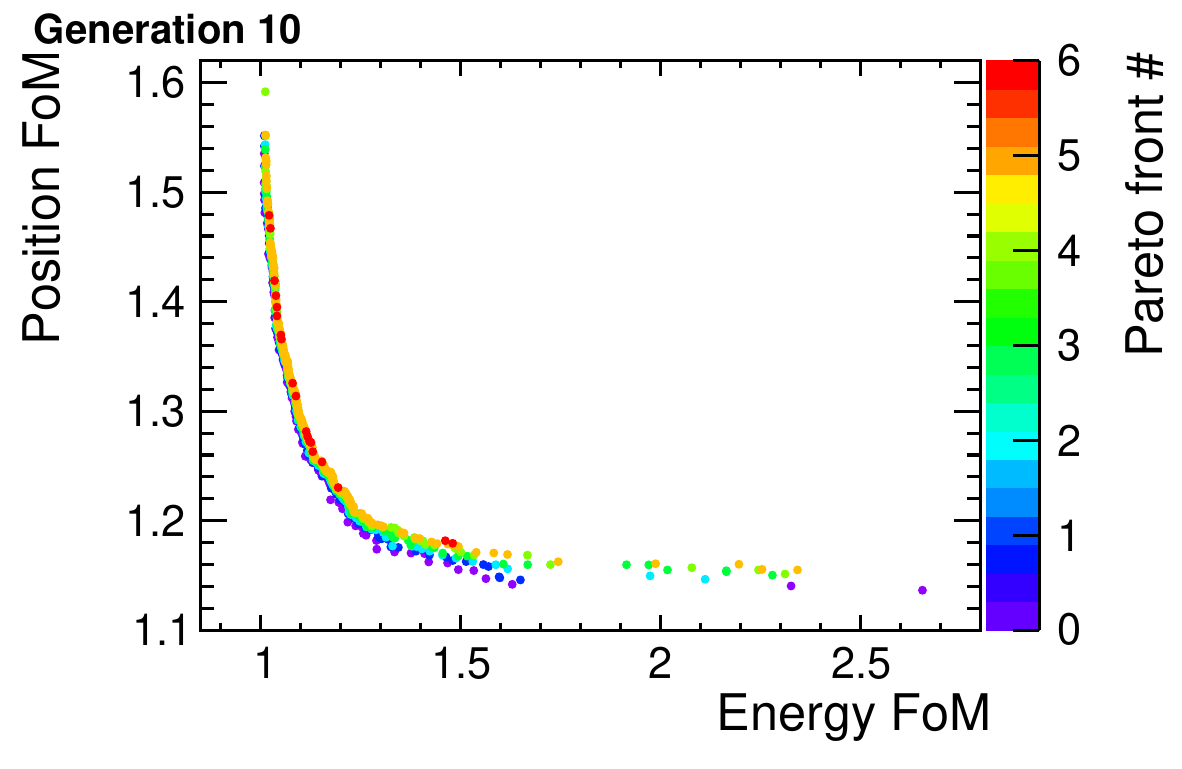}
\includegraphics[width=\twofigwidth]{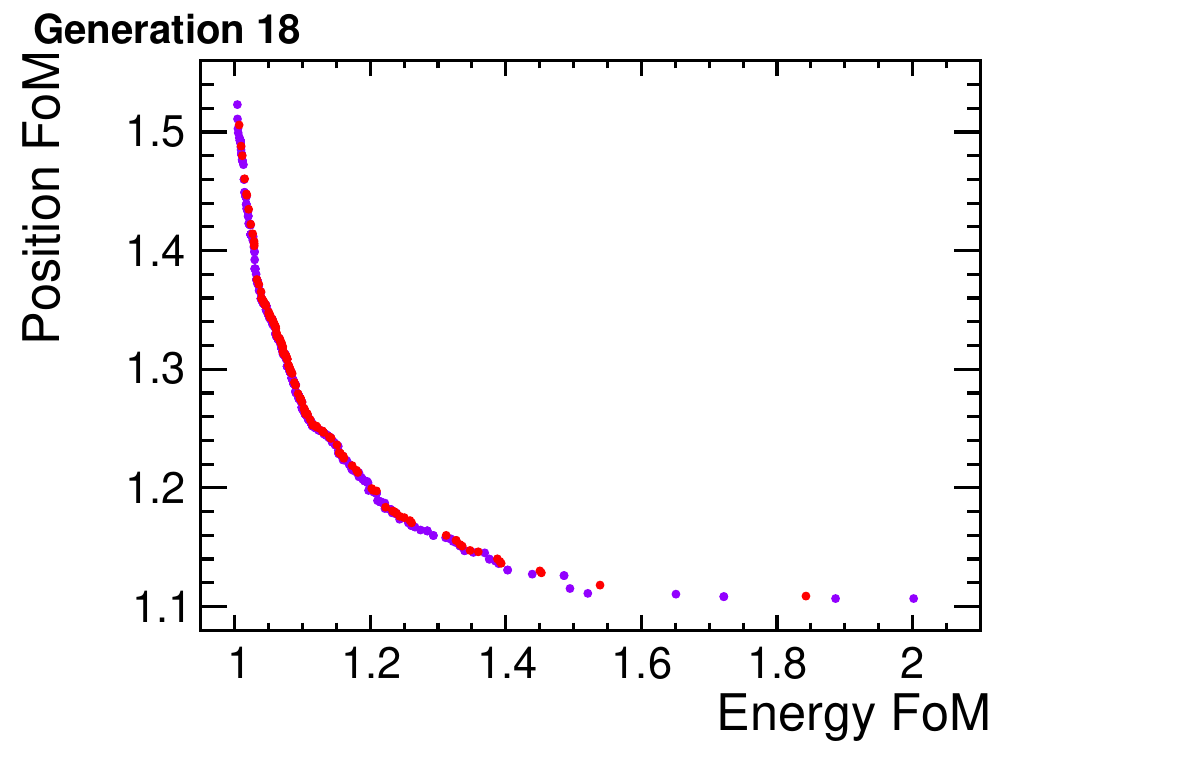}
\caption{Energy and position FoMs for the best 1000 layer
  configurations from subsequent configuration
  generations. For each generation, configurations are selected based
  on Pareto fronts, as indicated in the plot.}
\label{fig:pareto}
\end{figure}

Results of the genetic algorithm optimization of the longitudinal
calorimeter structure with non dominated sorting are presented in figure~\ref{fig:pareto}.
Presented are the best 1000 layer configurations from subsequent
generations, grouped in Pareto fronts (indicated by colour).
In the initial steps, about 20 Pareto fronts contribute to the best
1000 configurations (used as parents for the next generation), but the
front multiplicity is increasing fast.
After 18 generations 97\% of best configurations belong to the first
front and no further improvement is obtained.
It is important to notice that the final selection of configurations
forms a curve in the FoM plane tangential to the
envelope of the targeted optimization results presented in figure~\ref{fig:multi}.

\begin{figure}[tb]
\centering
\includegraphics[height=6cm]{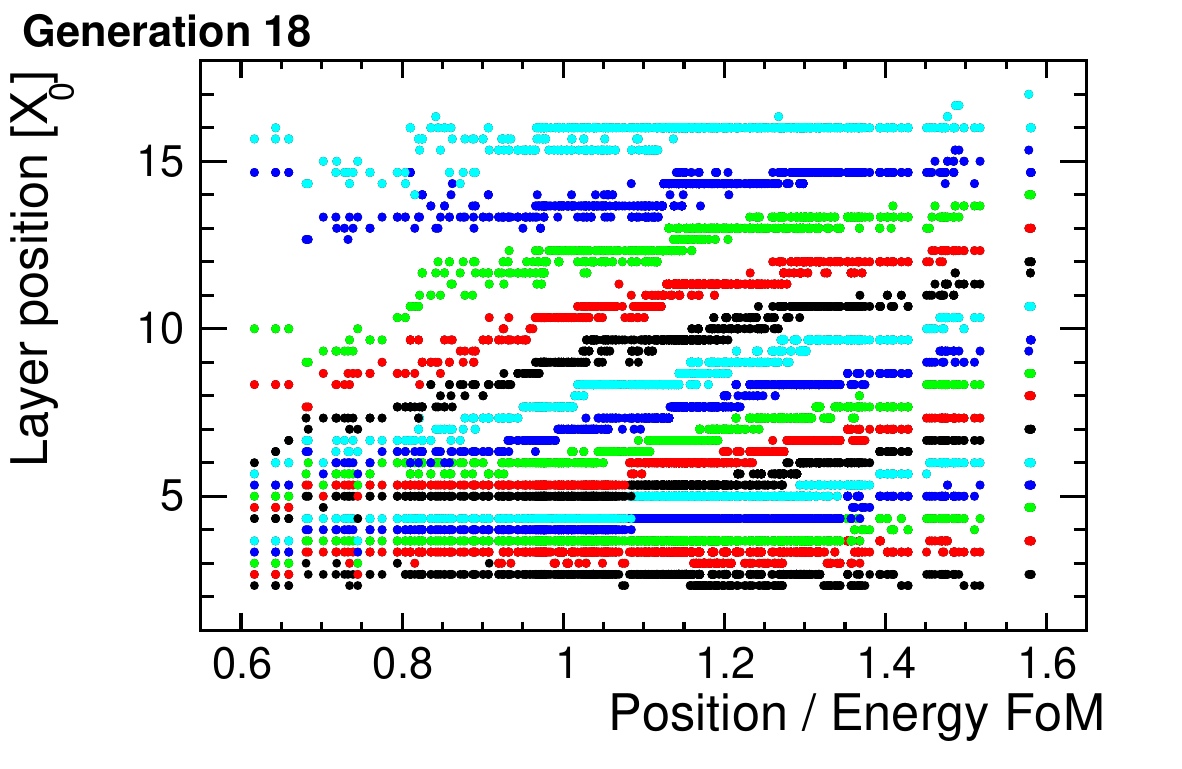}
\caption{Optimal positions of the calorimeter layers as a function of
  the position to energy FoM ratio defined as the
  optimization goal. Presented are the best 1000 configurations from
  the multi-objective optimization based on the Pareto fronts.}
\label{fig:pareto2}
\end{figure}

The procedure presented here does not allow for the final selection of
the best calorimeter configuration.
However, a collection of quasi-optimal configuration options can be
selected with the proposed procedure, which can be further studied in more
details.
The selected set of configurations will include very different
configuration options, but when a particular optimization goal is
considered, an optimal configuration should be found within this set.
This diversity is also illustrated in figure~\ref{fig:pareto2},
where positions of the calorimeter layers for the best configurations
are plotted as a function of the position to energy resolution FoMs.
One can clearly see that configurations with most of the instrumented
layers in the first 7\,X$_{0}$ of the calorimeter are preferred for
measurements dominated by position resolution while more uniform
instrumentation is required when energy resolution is more important.

\section{Conclusions}

Calorimeters are central components for most modern particle physics
experiments used primarily for the detection of charged and neutral
particles, as well as for reconstruction of their energy, position,
direction and (partial) identification.
Substantial progress in the development of new sensor technologies opened new avenues for
compact sampling calorimeter design.
This also opens a new field for the optimization of the calorimeter
segmentation and readout structure for the best performance of the device.

Presented in this contribution are optimization studies which were
performed in the framework of the LUXE ECAL-P group.
A general, semi-analytical framework for calorimeter response calibration and optimization,
including response linearity, energy resolution and position
resolution goals, has been developed allowing for very efficient
comparison of different calorimeter configurations.
When extended to the more general case of a high density electromagnetic calorimeter,
genetic algorithm looks like an efficient tool for finding the optimal calorimeter configuration. 
While optimization results strongly depend on the optimization goal selected,
non dominated sorting based on Pareto frontiers can be used to find a
larger set of optimal configuration, which can then be considered in
more details, for a particular measurement. 
The approach presented is very general and it can be used also for other experiments
and calorimeter concepts. 

\vspace*{0.5cm}

\begin{acknowledgement}

This study was performed in the framework of the ECAL group of the
LUXE collaboration. We wish to thank all group members who
contributed to the discussions of the results and the manuscript.
We would also like to acknowledge support and hospitality of WIS,
Rehovot and IFIC, Valencia during our visits.

\end{acknowledgement}


\bibliography{zarnecki_optimization_lcws2024_proc.bib} 

\end{document}